\documentclass[twocolumn]{aastex62}
\usepackage{color}
\usepackage{threeparttable}

\shortauthors{Kuriki, Sano, Kuno et al.}

\received{November 22, 2017}
\revised{July 5, 2018}
\accepted{July 31, 2018}
\begin{document}
\title{Discovery of molecular and atomic clouds associated with the gamma-ray supernova remnant Kesteven~79}

%\correspondingauthor{H. Sano}
%\email{sano@a.phys.nagoya-u.ac.jp}

\author{M. Kuriki}
\affiliation{Department of Physics, Graduate School of Pure and Applied Sciences, University of Tsukuba, 1-1-1 Tennodai, Tsukuba 305-8577, Japan}

\author{H. Sano}
\affiliation{Institute for Advanced Research, Nagoya University, Furo-cho, Chikusa-ku, Nagoya 464-8601, Japan; sano@a.phys.nagoya-u.ac.jp}
\affiliation{Department of Physics, Nagoya University, Furo-cho, Chikusa-ku, Nagoya 464-8601, Japan}

\author{N. Kuno}
\affiliation{Department of Physics, Graduate School of Pure and Applied Sciences, University of Tsukuba, 1-1-1 Tennodai, Tsukuba 305-8577, Japan}
\affiliation{Center for Integrated Research in Fundamental Science and Technology (CiRfSE), University of Tsukuba, 1-1-1 Tennodai, Tsukuba 305-8571, Japan}

\author{M. Seta}
\affiliation{Department of Physics, School of Science and Technology, Kwansei Gakuin University, 2-1 Gakuen, Sanda 669-1337, Japan}

\author{Y. Yamane}
\affiliation{Department of Physics, Nagoya University, Furo-cho, Chikusa-ku, Nagoya 464-8601, Japan}

\author{T. Inaba}
\affiliation{Department of Physics, Nagoya University, Furo-cho, Chikusa-ku, Nagoya 464-8601, Japan}

\author{T. Nagaya}
\affiliation{Department of Physics, Nagoya University, Furo-cho, Chikusa-ku, Nagoya 464-8601, Japan}

\author{S. Yoshiike}
\affiliation{Department of Physics, Nagoya University, Furo-cho, Chikusa-ku, Nagoya 464-8601, Japan}

\author{K. Okawa}
\affiliation{Department of Physics, Nagoya University, Furo-cho, Chikusa-ku, Nagoya 464-8601, Japan}

\author{D. Tsutsumi}
\affiliation{Department of Physics, Nagoya University, Furo-cho, Chikusa-ku, Nagoya 464-8601, Japan}

\author{Y. Hattori}
\affiliation{Department of Physics, Nagoya University, Furo-cho, Chikusa-ku, Nagoya 464-8601, Japan}

\author{M. Kohno}
\affiliation{Department of Physics, Nagoya University, Furo-cho, Chikusa-ku, Nagoya 464-8601, Japan}

\author{S. Fujita}
\affiliation{Department of Physics, Nagoya University, Furo-cho, Chikusa-ku, Nagoya 464-8601, Japan}

\author{A. Nishimura}
\affiliation{Department of Physics, Nagoya University, Furo-cho, Chikusa-ku, Nagoya 464-8601, Japan}

\author{A. Ohama}
\affiliation{Department of Physics, Nagoya University, Furo-cho, Chikusa-ku, Nagoya 464-8601, Japan}

\author{M. Matsuo}
\affiliation{Nobeyama Radio Observatory, National Astronomical Observatory of Japan, National Institutes of Natural Sciences, 462-2 Nobeyama, Minamimaki, Minamisaku-gun 384-1305, Japan}
\affiliation{Graduate Schools of Science and Engineering, Kagoshima University, 1-21-35 Korimoto, Kagoshima 890-0065, Japan}

\author{Y. Tsuda}
\affiliation{Meisei University, 2-1-1 Hodokubo, Hino 191-0042, Japan}

\author{K. Torii}
\affiliation{Nobeyama Radio Observatory, National Astronomical Observatory of Japan, National Institutes of Natural Sciences, 462-2 Nobeyama, Minamimaki, Minamisaku-gun 384-1305, Japan}

\author{T. Minamidani}
\affiliation{Nobeyama Radio Observatory, National Astronomical Observatory of Japan, National Institutes of Natural Sciences, 462-2 Nobeyama, Minamimaki, Minamisaku-gun 384-1305, Japan}

\author{ T. Umemoto}
\affiliation{Nobeyama Radio Observatory, National Astronomical Observatory of Japan, National Institutes of Natural Sciences, 462-2 Nobeyama, Minamimaki, Minamisaku-gun 384-1305, Japan}

\author{G. Rowell}
\affiliation{School of Physical Sciences, University of Adelaide, North Terrace, Adelaide, SA 5005, Australia}

\author{A. Bamba}
\affiliation{Department of Physics, The University of Tokyo, 7-3-1 Hongo, Bunkyo-ku, Tokyo 113-0033, Japan}
\affiliation{Research Center for the Early Universe, School of Science, The University of Tokyo, 7-3-1 Hongo, Bunkyo-ku, Tokyo 113-0033, Japan}

\author{K. Tachihara}
\affiliation{Department of Physics, Nagoya University, Furo-cho, Chikusa-ku, Nagoya 464-8601, Japan}

\author{Y. Fukui}
\affiliation{Institute for Advanced Research, Nagoya University, Furo-cho, Chikusa-ku, Nagoya 464-8601, Japan; sano@a.phys.nagoya-u.ac.jp}
\affiliation{Department of Physics, Nagoya University, Furo-cho, Chikusa-ku, Nagoya 464-8601, Japan}

\begin{abstract}
We carried out $^{12}$CO($J$ = 1--0) observations of the Galactic gamma-ray supernova remnant (SNR) Kesteven~79 using the Nobeyama Radio Observatory 45 m radio telescope, which has an angular resolution of $\sim20$ arcsec. We identified molecular and atomic gas interacting with Kesteven~79 whose radial velocity is $\sim80$ km s$^{-1}$. The interacting molecular and atomic gases show good spatial correspondence with the X-ray and radio shells, which have an expanding motion with an expanding velocity of $\sim4$ km s$^{-1}$. The molecular gas associated with the radio and X-ray peaks also exhibits a high-intensity ratio of CO 3--2/1--0 $>$ 0.8, suggesting a kinematic temperature of $\sim24$ K, owing to heating by the supernova shock. We determined the kinematic distance to the SNR to be $\sim5.5$ kpc and the radius of the SNR to be $\sim8$ pc. The average interstellar proton density inside of the SNR is $\sim360$ cm$^{-3}$, of which atomic protons comprise only $\sim10$ $\%$. Assuming a hadronic origin for the gamma-ray emission, the total cosmic-ray proton energy above 1 GeV is estimated to be $\sim5 \times 10^{48}$ erg. 
\end{abstract}
\keywords{cosmic rays --- ISM: clouds --- ISM: molecules --- gamma rays: ISM --- ISM: supernova remnants --- X-rays: individual objects (Kesteven~79, G33.6$+$0.1, 4C00.70, HC13)}

\section{Introduction} \label{sec:intro}
It has been a longstanding matter of debate as to how cosmic-ray (CR) protons, which are the major components of CRs, are accelerated in the Galaxy. Supernova remnants (SNRs) are the most likely candidates for the acceleration sites of Galactic CRs, the energies of which extend up to $\sim3 \times 10^{15}$ eV, the so-called ``knee'' energy \citep[e.g.,][]{1964SvA.....8..342G,1952PThPh...8..571H}. Theoretical studies have predicted that CRs can be efficiently accelerated in SNR shockwaves via diffusive shock acceleration \citep[DSA; e.g.,][]{1978MNRAS.182..147B,1978ApJ...221L..29B}. However, the principal acceleration sites of CR protons still remain elusive because of the paucity of observational evidence.

Gamma-ray bright SNRs hold a key to understanding the origin of CRs. Gamma-rays from SNRs are generally produced by CR protons and electrons through the hadronic or leptonic processes. In hadronic process, interactions between CR protons and interstellar protons create a neutral pion which decays into two gamma-ray photons. On the other hand, CR electrons energize interstellar low-energy photon into gamma-ray via the inverse Compton effect. Moreover, leptonic gamma-rays can be also produced by the non-thermal Bremsstrahlung of CR electrons. In general, it is very difficult to distinguish the hadronic and leptonic processes by the spectral modeling alone \citep[e.g.,][]{2012ApJ...744...71I}.

Recently, investigations of the interstellar medium associated with gamma-ray SNRs have received much attention as a tool for understanding the origins of CRs. \cite{2012ApJ...746...82F} demonstrated good spatial correspondence between the total column density of interstellar protons (both atomic and molecular components) and TeV gamma rays in the young ($\sim2,000$ yr) SNR RX~J1713.7$-$3946. This provides one of the essential conditions for generating hadronic gamma rays via neutral pion decay associated with proton-proton interactions. On the basis of an estimated mean interstellar proton density of $\sim130$ cm$^{-3}$, they concluded that CR protons are accelerated in the SNR and that the total energy in CRs is $\sim10^{48}$ erg, which corresponds to $\sim0.1$ $\%$ of the total $\sim10^{51}$ erg kinematic energy released in the supernova explosion. Subsequent studies showed similar results for other young TeV gamma-ray SNRs (e.g., HESS~J1731$-$347, \citealt{2014ApJ...788...94F}; Vela Jr., \citealt{2017ApJ...850...71F}). On the other hand, middle aged SNRs ($\sim20,000$ yr), which are bright in GeV gamma rays, have a total CR proton energy of the order of $10^{49}$--$10^{50}$ erg, based on a mean interstellar proton density of a few hundred to less than 1,000 cm$^{-3}$ (e.g., W44, \citealt{2013ApJ...768..179Y}). To better understand the origin of CR protons and their energies, we need to have a larger sample of hadronic gamma rays from the SNRs.

\begin{figure}
\begin{center}
\includegraphics[width=\linewidth,clip]{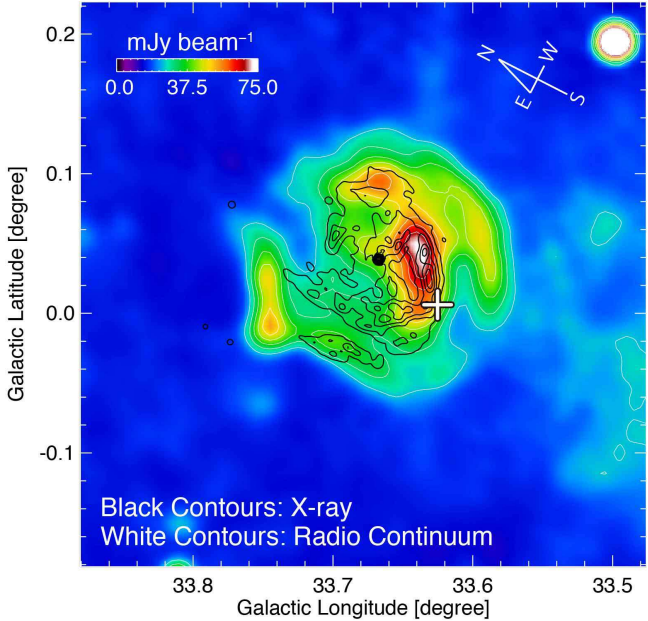}
\caption{Distribution of the VLA radio continuum at 1.4 GHz superposed on the $Chandra$ X-ray contours in the energy band of 0.3--10.0 keV \citep{2006AJ....132.1158S,2004ApJ...605..742S}. The black and white contours indicate the X-ray and radio continuum, respectively. The contour level is every 0.3 $\times$ 10$^{-6}$ counts pixel$^{-1}$ s$^{-1}$ from 0.2 $\times$ 10$^{-6}$ pixel$^{-1}$ s$^{-1}$ for the X-rays; every 6 mJy beam$^{-1}$ from 25 mJy beam$^{-1}$ ($>$ 45 $\sigma$) for the radio continuum. The position of GeV gamma-ray peak is indicated by the yellow cross \citep{2014ApJ...783...32A}. The point spread function of gamma-rays is $\sim$ 0\fdg3 (FWHM).}
\label{fig1}
\end{center}
\end{figure}

Kesteven~79 (hereafter Kes~79; also known as G33.6$+$0.1, 4C00.70, or HC13) is a Galactic SNR located at ($l$, $b$) $\sim$ (33\fdg6, 0\fdg1). It has been well studied at radio continuum and thermal X-ray wavelengths \citep{1975AuJPA..37...39C,1981MNRAS.195...89C,1975AJ.....80..437D,1975AJ.....80..679B,1977A&A....55...11A,1989ApJ...336..854F,1991AJ....102..676V,1992AJ....103..943K,1995ApJ...439..715S,2002PASJ...54..735T,2004ApJ...605..742S,2009A&A...507..841G,2014ApJ...783...32A,2016PASJ...68S...8S,2016ApJ...831..192Z}. Figure \ref{fig1} shows the Very Large Array (VLA) radio-continuum image superposed on $Chandra$ X-ray contours for the energy band 0.3--10.0 keV. The X-ray emission is bright inside the radio-continuum shell, which has a diameter of $\sim10$ arcmin, indicating that it is a mixed-morphology SNR, according to \cite{1998ApJ...503L.167R}. The radio-continuum shell appears to consist of two incomplete, concentric shells \citep[e.g.,][]{1991AJ....102..676V}. \cite{2004ApJ...605..742S} found that the multiple filaments and the ``protrusion'' in thermal X-rays obtained by $Chandra$ show good spatial correspondence with the radio-continuum shell \cite[e.g.,][]{2009A&A...507..841G}. \cite{2016PASJ...68S...8S} presented the most reliable X-ray spectroscopy for Kes~79 using $Suzaku$ with a low and stable non-X-ray background and a good spectroscopic resolution. They discovered that the X-rays from this SNR are consistent with a two-temperature model: a collisional ionization equilibrium (CIE) plasma (with $k{T_\mathrm{e}} \sim 0.2$ keV) and a non-equilibrium ionized (NEI) plasma (with $k{T_\mathrm{e}} \sim 0.8$ keV). The abundance ratios and the mass of the ejecta are consistent with a core-collapse SNR originating from a progenitor of mass $\sim30$--40 $M_\odot$. The dynamical age of Kes~79 was also estimated to be $\sim2.7 \times 10^4$ yr, assuming a distance of $\sim7$ kpc and the Sedov self-similar solution \citep{1959sdmm.book.....S}.

Kes~79 is also known to be associated with a dense molecular cloud. \cite{1992MNRAS.254..686G} used the $^{12}$CO($J$ = 1--0) and HCO$^{+}$($J$ = 1--0) emission lines obtained with the National Radio Astronomy Observatory 12 m radio telescope to show that a dense molecular cloud with $V_\mathrm{LSR} \sim 105$ km s$^{-1}$ exhibits good spatial correspondence with the radio shell of Kes~79. They claimed that the SNR shockwaves interact with this molecular cloud, the kinematic distance to which is $\sim7.1$ kpc. Subsequently, \cite{2016ApJ...831..192Z} confirmed this using the CO($J$ = 1--0, 2--1, 3--2) emission lines obtained with the 13.7 m millimeter-wavelength telescope of the Purple Mountain Observatory, K\"{o}lner Observatory for Submillimeter Astronomy, and archival datasets of the James Clerk Maxwell Telescope (JCMT). They found a broad molecular line spanning a velocity range of 20 km s$^{-1}$ toward the eastern X-ray/radio filaments, indicating that the occurrence of shock-interactions. \cite{2016ApJ...816....1K} also pointed out a broad molecular line using $^{12}$CO($J$ = 2--1) data from the 10 m Submillimeter Telescope. These previous studies showed that the shock-cloud interaction is a feasible scenario to understand qualitatively the physical association between the SNR shells and the associated ISM. To obtain a conclusive piece of evidence for the shock-cloud interaction, we need more detailed quantitative studies of the interstellar medium toward SNRs as follows: (1) shock-excited OH masers \citep[e.g., W44 and W28,][]{1968ApJS...15..131G}, (2) interstellar expanding shells created by SNR shockwaves and/or stellar winds from the progenitor of the supernova explosion (e.g., RX~J1713.7$-$3946, \citealt{2012ApJ...746...82F}; RCW~86, \citealt{2017JHEAp..15....1S}), or (3) high-temperature gas from more intensive shock heating \citep[e.g., W44 and IC443,][]{1998ApJ...505..286S}. For Kes~79, such detailed studies using CO/H{\sc i} datasets are needed to clarify the velocity ranges of interstellar gas associated with the SNR.

Investigating the interstellar gas associated with the SNR Kes~79 is also important for understanding the origin of CR protons. Recently, \cite{2014ApJ...783...32A} discovered a GeV gamma-ray excess in the direction of Kes~79 by using the $Fermi$ Large Area Telescope (LAT). The region of gamma-ray excess is consistent with the position of the molecular cloud thought to be associated with the SNR (see yellow cross in Figure \ref{fig1}). The gamma-ray spectral index $2.62 \pm 0.12$ is well described by hadronic scenario, whereas lepton dominant scenarios failed to explain the gamma-ray spectrum with realistic parameters. In addition, there is no evidence in favor of strong gamma-ray contribution from a pulsar wind nebula \citep{2014ApJ...783...32A}. Furthermore, \cite{2016PASJ...68S...8S} discovered a K-shell line of Fe at 6.4 keV toward the east of Kes~79. They suggest that this line originates from neutral Fe in the molecular cloud within which the SNR is embedded, which is ionized by locally accelerated CR protons with energies of the order of MeV.

In the present paper, we aim to identify the interstellar molecular and atomic gas associated with Kes~79 to determine the CR proton energies in the SNR using new CO($J$ = 1--0) datasets obtained with the Nobeyama Radio Observatory (NRO) 45 m telescope using new receiver named the FOur-beam REceiver System on the 45 m Telescope \citep[FOREST;][]{2016SPIE.9914E..1ZM} and from archived H{\sc i}/CO datasets. This paper is organized as follows. In Section \ref{sec:obs}, we describe the CO, X-ray, H{\sc i}, and radio-continuum datasets. Section \ref{sec:results} consists of four subsections: Subsection \ref{subsec:overview} gives the distributions of CO, X-rays, and the radio continuum in the direction of Kes~79; Subsection \ref{subsec:expanding} gives the expanding shell structures of CO and H{\sc i}; Subsection \ref{subsec:ratio} presents the maps of the CO $J$ = 3--2/1--0 intensity ratio; and Subsection \ref{subsec:lvg} gives the physical conditions in the molecular cloud. The discussion and summary are given in Sections \ref{sec:discussion} and \ref{sec:summary}, respectively. 

\section{OBSERVATIONS AND DATA REDUCTION} \label{sec:obs}
\subsection{CO} \label{sec:obs.co}
We performed $^{12}$CO($J$ = 1--0) observations with the NRO 45 m telescope as part of the FOREST Unbiased Galactic Plane Imaging survey with Nobeyama 45 m telescope \citep[FUGIN;][]{2017PASJ...69...78U}. The survey was carried out from 2014 to 2017 using the on-the-fly mapping mode \citep{2008PASJ...60..445S}. It covered a Galactic longitude range from 10$^{\circ}$ to 50$^{\circ}$ and a Galactic latitude range of within $\pm1^{\circ}$ in the inner region of our Galaxy. Observations of Kes~79 were conducted in June 2016. The frontend was FOREST, which includes the four beams with dual polarizations, and a two-sideband superconductor-insulator-superconductor receiver \citep{2016SPIE.9914E..1ZM}. The effective spatial resolution was 20 arcsec, at a frequency of 115 GHz. The typical system temperature was $\sim250$ K at 70 deg, including the atmosphere. The spectrometer was equipped with the Spectral Analysis Machine for the 45 m telescope \citep[SAM45;][]{2011URSI..30...2K,2012PASJ...64...29K}, which processes 16 intermediate frequency (IF) signals and outputs 4,096 channels per IF, each of 1~GHz bandwidth. The velocity and frequency resolutions were 1.3 km s$^{-1}$ ch$^{-1}$ and 244.14 kHz, respectively. The pointing accuracy was checked every hour by observing SiO maser sources and we achieved the pointing error within $\sim2$ arcsec. The main beam efficiency $\eta_\mathrm{mb}$ of 0.43 $\pm$ 0.02 was applied to convert from the antenna temperature $T_\mathrm{A}^{*}$ to the brightness temperature $T_\mathrm{mb}$. We observed standard sources W51 [$\alpha_\mathrm{J2000}$ = $19^{\mathrm{h}}23^{\mathrm{m}}50\fs0$, $\delta_\mathrm{J2000}$ = $14{^\circ}06\arcmin00\farcs0$] \citep{2014BASI...42...47G} and Orion KL [$\alpha_\mathrm{J2000}$ = $05^{\mathrm{h}}35^{\mathrm{m}}14\fs16$, $\delta_\mathrm{J2000}$ = $-05{^\circ}22\arcmin21\farcs5$] \citep{2003AJ....126.1423K} to verify the performance of the telescope. The map we use in this paper covers the region of $l$ = 33$^{\circ}$--34$^{\circ}$ and $|b| \leq 1^{\circ}$ (1$^{\circ} \times 2^{\circ}$), and we reprocessed the data using the NOSTAR software provided by the NRO. The final dataset was smoothed to a pixel size of 30 arcsec, with a typical rms noise level of $\sim0.46$ K at a velocity resolution of 1 km s$^{-1}$. 

To measure the intensity ratio of $^{12}$CO $J$ = 3--2/1--0, we also used the $^{12}$CO($J$ = 3--2) data from the CO High-Resolution Survey \citep[COHRS;][]{2013ApJS..209....8D} obtained with the JCMT. We smoothed the data to a spatial resolution of 20 arcsec, and a velocity resolution of 1 km s$^{-1}$. We applied the main beam efficiency $\eta_\mathrm{mb}$ of 0.61 to convert from the antenna temperature $T_\mathrm{A}^{*}$ to the brightness temperature $T_\mathrm{mb}$. The final rms noise level was $\sim0.32$ K, at a velocity resolution of 1 km s$^{-1}$.

\subsection{H{\sc i}} \label{sec:obs.hi}
The H{\sc i} and 21 cm radio-continuum data are from the VLA Galactic Plane Survey \citep[VGPS;][]{2006AJ....132.1158S}. The spatial and velocity resolutions for H{\sc i} are $\sim1$ arcmin and $\sim1.56$ km s$^{-1}$, respectively, and the rms noise is 2 K ch$^{-1}$. The continuum image without H{\sc i} lines has a spatial resolution of 1 arcmin and a typical rms noise level of $\sim0.3$ K.

\subsection{X-rays} \label{sec:obs.x}
We used X-ray data taken with $Chandra$ from July to August 2001, for which the observation ID is 1982 \citep[PI: Seward,][]{2004ApJ...605..742S}. We used the CIAO software version 4.8.1 with CALDB version 4.7.1 for data reduction and imaging analysis \citep{2006SPIE.6270E..1VF}. The dataset was reprocessed using the script ``chandra repro.'' Combined, energy-filtered, exposure-corrected, and binned images were produced using the scripts ``fluximage'' and ``dmcopy'' specifying the energy-band of 0.3--10.0 keV. The total exposure time and effective exposure time were 29.95 and 29.27 ks, respectively.

\section{RESULTS} \label{sec:results}
\subsection{Overview of CO, X-ray, and radio-continuum distributions} \label{subsec:overview}
First, we searched for a good spatial correspondence between the X-ray/radio-continuum and CO/H{\sc i} intensities over the velocity range from $-20$ to 120 km s$^{-1}$ to identify which interstellar gases are associated with the SNR. The method has been applied in several previous investigations \citep[e.g.,][]{2005ApJ...631..947M,2012ApJ...746...82F,2017JHEAp..15....1S}. We found possible counterparts to the interstellar molecular and atomic gas to lie in the velocity ranges of $V_\mathrm{LSR}$ = 30.5--39.5 km s$^{-1}$ (hereafter referred to as the ``30 km s$^{-1}$ cloud''), $V_\mathrm{LSR}$ = 82.5--86.5 km s$^{-1}$ (hereafter referred to as the ``80 km s$^{-1}$ cloud''), and $V_\mathrm{LSR}$ = 99.5--112.5 km s$^{-1}$ (hereafter referred to as the ``100 km s$^{-1}$ cloud'').

\begin{figure*}
\begin{center}
\includegraphics[width=\linewidth,clip]{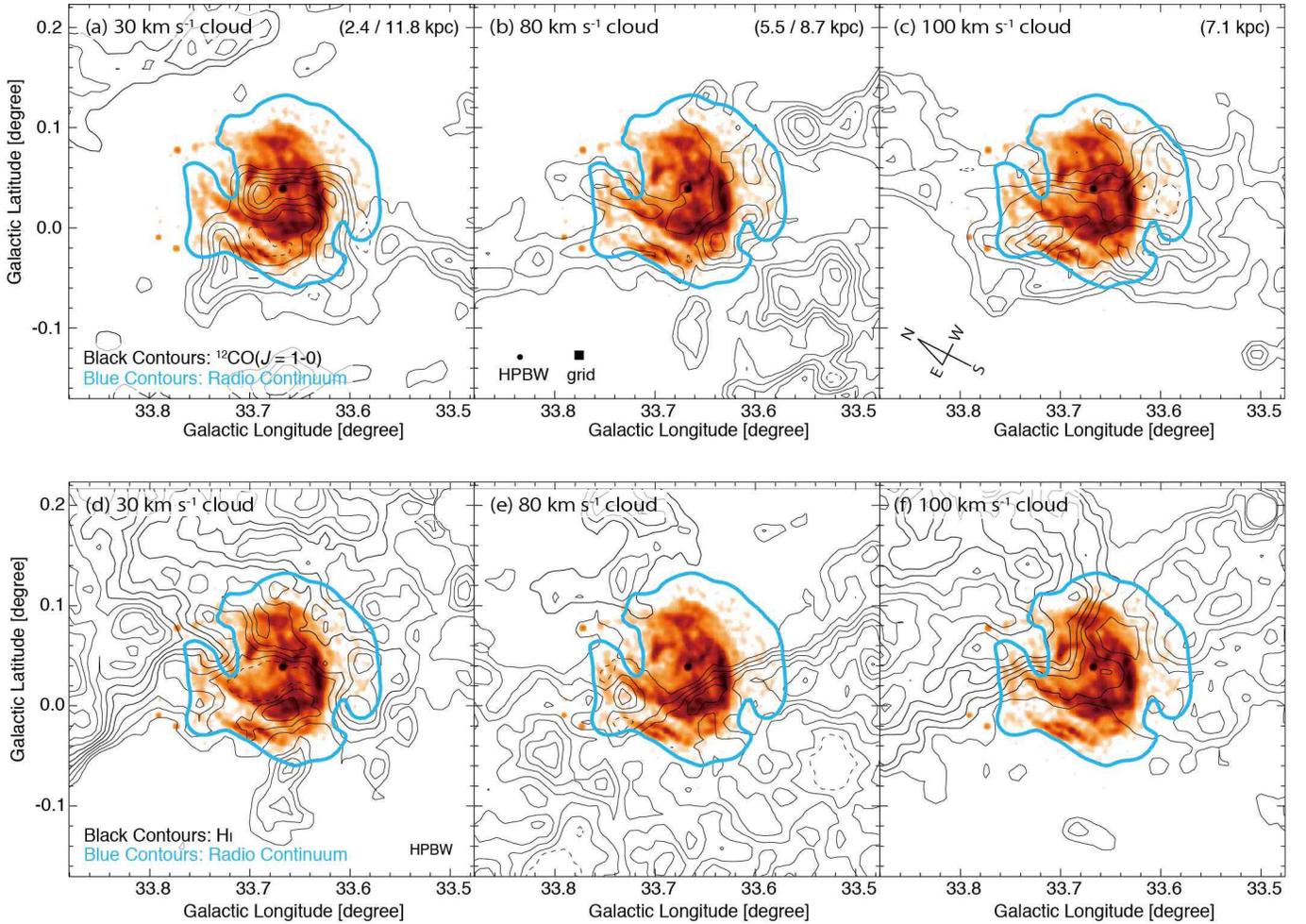}
\caption{Integrated intensity distributions of $^{12}$CO($J$ = 1--0) (a--c, $black$ $contours$) and H{\sc i} (d--f, $black$ $contours$) superposed with the X-ray image as shown in Figure \ref{fig1}. The velocity range is $V_\mathrm{LSR}$ = 30.5--39.5 km s$^{-1}$ for (a) and (d); $V_\mathrm{LSR}$ = 82.5--86.5 km s$^{-1}$ for (b) and (e); $V_\mathrm{LSR}$ = 99.5--112.5 km s$^{-1}$ for (c) and (f). The contour levels are every 3 K km s$^{-1}$ from 10 K km s$^{-1}$ for (a); every 4 K km s$^{-1}$ from 18 K km s$^{-1}$ for (b); every 10 K km s$^{-1}$ from 60 K km s$^{-1}$ for (c); every 30 K km s$^{-1}$ from 840 K km s$^{-1}$ for(d); every 20 K km s$^{-1}$ from 380 K km s$^{-1}$ for (e); every 40 K km s$^{-1}$ from 690 K km s$^{-1}$ for (f). The blue contours indicate the boundary of radio continuum. The contour level is 25 mJy beam$^{-1}$ ($>$ 45 $\sigma$). The kinematic distance corresponding to the velocity range are indicated in the top right of each panel.}
\label{fig2}
%\vspace*{0.5cm}
\end{center}
\end{figure*}%

Figures \ref{fig2}a, \ref{fig2}b, and \ref{fig2}c show the $^{12}$CO($J$ = 1--0) intensity contours of the 30, 80, and 100 km s$^{-1}$ clouds superposed on the X-ray image and the radio-continuum boundaries. In Figure \ref{fig2}a, the 30 km s$^{-1}$ cloud has a shell-like structure near Kes~79. The molecular cloud appears along the X-ray shell and the radio continuum boundary from east to south. By contrast, the northwestern CO cloud lies only across the center of Kes~79, and has no significant spatial correspondence with the X-ray shell. Furthermore, there is no counterpart to the molecular cloud toward the western shells of the X-ray and radio continuum. The kinematic distance to the 30 km s$^{-1}$ cloud was estimated to be 2.4 kpc (near side) and 11.8 kpc (far side) by adopting the Galactic rotation curve model of \cite{1993A&A...275...67B}.

In Figure \ref{fig2}b, the overall distribution of the 80 km s$^{-1}$ cloud tends to encircle the X-ray shell-like structure. In particular, the southern rim of X-rays show a good spatial anti-correlation with the CO peaks at ($l$, $b$) $\sim$ (33\fdg64, $-0\fdg2$) and (33\fdg61, 0\fdg01). In addition, two CO filamentary structures at ($l$, $b$) $\sim$ (33\fdg64, $-0\fdg2$) and (33\fdg61, 0\fdg01) are situated along the outer rims of the X-ray double-shell-like structures. The 80 km s$^{-1}$ cloud is also embedded along the radio-continuum boundary. The kinematic distance of the 80 km s$^{-1}$ cloud was estimated to be 5.5 kpc (near side) or 8.7 kpc (far side).

\begin{figure*}
\begin{center}
\includegraphics[width=\linewidth,clip]{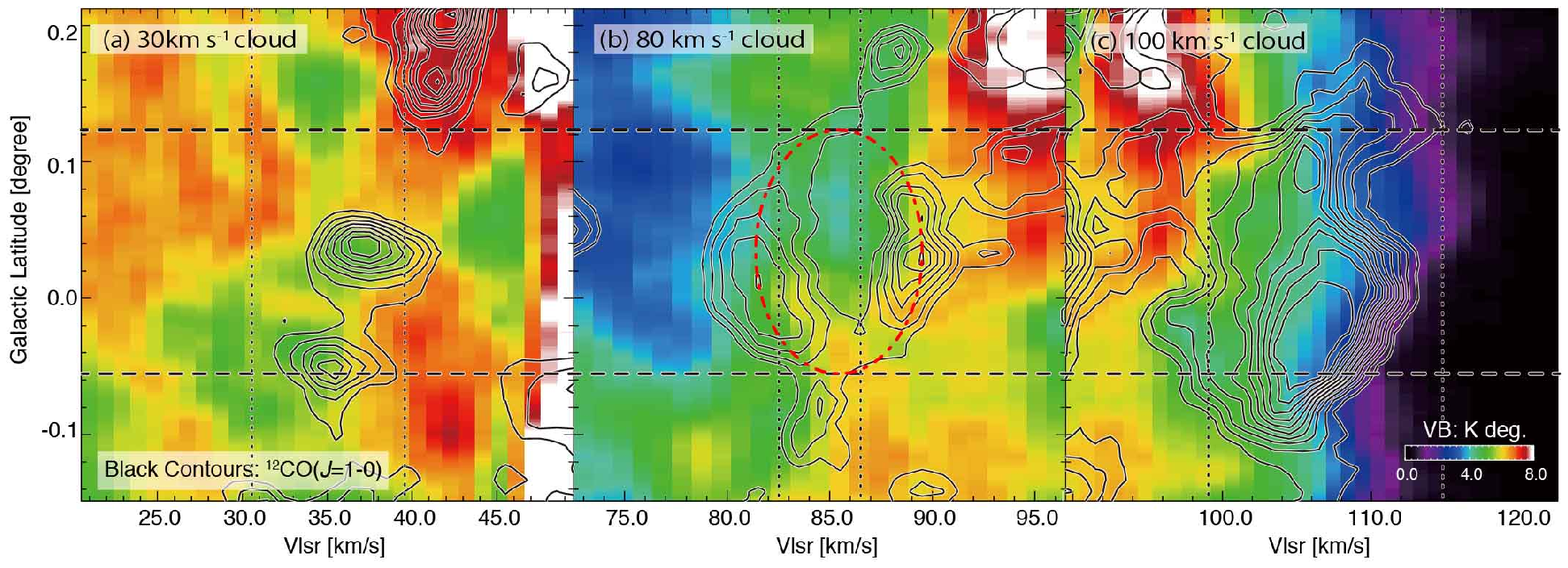}
\caption{Velocity--Galactic latitude diagrams of H{\sc i} superposed on the $^{12}$CO($J$ = 1--0) contours in (a) the 30 km s$^{-1}$ cloud, (b) the 80 km s$^{-1}$ cloud, and (c) the 100 km s$^{-1}$ cloud. The integrated range of Galactic longitude is from 33\fdg64 to 33\fdg71. The CO contour levels are every 0.06 K degree from 0.1 K degree for (a) and (c); every 0.05 K degree from 0.22 K degree for (b). The vertical and horizontal dashed lines indicate the integrated velocity range and the size of the radio continuum shell, respectively.}
\label{fig3}
\end{center}
\end{figure*}%

\begin{figure*}
\begin{center}
\includegraphics[width=\linewidth,clip]{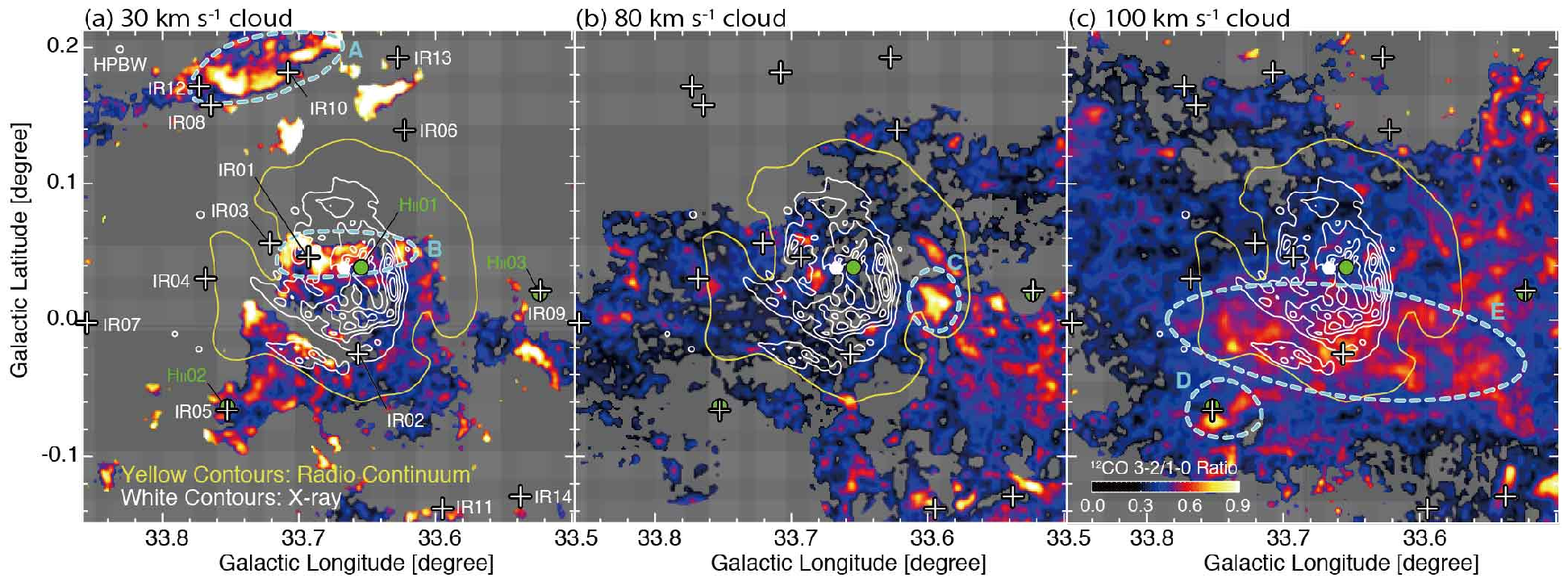}
\caption{Spatial distribution of the line intensity ratio between $^{12}$CO($J$ = 3--2) and $^{12}$CO($J$ = 1--0). The velocity range is $V_\mathrm{LSR}$ = 30.5--39.5 km s$^{-1}$ for (a); $V_\mathrm{LSR}$ = 82.5--86.5 km s$^{-1}$ for (b); $V_\mathrm{LSR}$ = 99.5--112.5 km s$^{-1}$ for (c). The superposed contours indicate the X-rays ($white$) and radio continuum ($yellow$). The contour levels of X-ray and radio continuum are the same as those in Figures \ref{fig1} and \ref{fig2}, respectively. The IRAS point sources and H{\sc ii} regions are indicated by the crosses and circles (see also Table \ref{table1}). The region bright in the intensity ratio of CO $J$ = 3--2/1--0 are enclosed by dashed lines.}
\label{fig4}
\end{center}
\vspace*{0.5cm}
\end{figure*}%

In Figure \ref{fig2}c, the 100 km s$^{-1}$ cloud has a strong CO peak located at ($l$, $b$) $\sim$ (33\fdg73, $-0\fdg2$), which corresponds to the region with the X-ray double-shell-like structures mentioned above. The CO cloud is also embedded along with the X-ray and radio boundaries from northeast to southwest. The 100 km s$^{-1}$ cloud has been considered to be associated with the SNR Kes~79 \citep{1992MNRAS.254..686G,2016ApJ...831..192Z}, and the kinematic distance is 7.1 kpc. We note that the CO intensity of the 100 km s$^{-1}$ cloud is 3--4 times larger than that of the 30 and 80 km s$^{-1}$ clouds, indicating that the 30 and 80 km s$^{-1}$ clouds are more diffuse than the 100 km s$^{-1}$ cloud. 

Figures \ref{fig2}d, \ref{fig2}e, and \ref{fig2}f show the same images as in Figures \ref{fig2}a, \ref{fig2}b, and \ref{fig2}c, but with the black contours now indicating the H{\sc i} intensities for each velocity component. For the 30 km s$^{-1}$ cloud, a shell-like structure of H{\sc i} agrees well with the X-ray shell. For the 80 km s$^{-1}$ cloud, the south and northeast parts of the H{\sc i} distribution are nicely associated with the X-ray filaments, whereas the western part of the X-ray shell has no H{\sc i} counterpart. For the 100 km s$^{-1}$ cloud, the H{\sc i} surrounds with only the northwestern part of the X-ray shell, which is complementary in spatial coverage to the CO distribution (see also Figure \ref{fig2}c). 

\subsection{Expanding shell structures of CO and H{\sc i}} \label{subsec:expanding}
Figures \ref{fig3}a, \ref{fig3}b, and \ref{fig3}c show the velocity-vs.-Galactic latitude diagrams for CO and H{\sc i} toward the 30, 80, and 100 km s$^{-1}$ clouds, respectively. The horizontal dashed lines indicate the apparent diameter of the radio continuum-boundary shown in Figure \ref{fig2}. The 30 and 100 km s$^{-1}$ clouds have no common structures in CO and H{\sc i}. In contrast, the 80 km s$^{-1}$ cloud shows a cavity-like structure, with intensity peaks at velocities of $\sim81.5$ and $\sim89.5$ km s$^{-1}$. The size of this cavity in Galactic latitude is roughly consistent with that of the radio-continuum boundary (diameter $\sim10$ arcmin); hence, this cavity is probable evidence for expanding gas motions, with an expansion velocity of $\sim4$ km s$^{-1}$.

\subsection{CO 3--2/1--0 ratio maps} \label{subsec:ratio}
Figures \ref{fig4}a, \ref{fig4}b, and \ref{fig4}c show the intensity ratio $^{12}$CO $J$ = 3--2/1--0 (hereafter $R_\mathrm{3-2/1-0}$) toward the 30, 80, and 100 km s$^{-1}$ clouds, respectively. We find five prominent regions A--E with high intensity ratios of $R_\mathrm{3-2/1-0} > 0.6$. Regions with $R_\mathrm{3-2/1-0} > 0.8$ are seen toward both the 30 km s$^{-1}$ (regions A and B) and 80 km s$^{-1}$ cloud (region C), and regions B and C are located within the radio shell boundary of the SNR. On the other hand, the 100 km s$^{-1}$ cloud has a region with $R_\mathrm{3-2/1-0} \sim 0.6$ elongated from the northeast to south of the SNR shell, with a size of 0.25 degrees. We also note that regions A, B, D, and E contain H{\sc ii} regions or IRAS point sources (see Table \ref{table1}), indicating the existence of stellar heating.

\subsection{Large velocity gradient analysis} \label{subsec:lvg}
We focus here on region C, which has no counterpart among the extra heating sources except for the SNR shockwaves. To investigate the physical properties of the molecular cloud in region C (hereafter referred to as ``cloud C''), we preformed a large velocity gradient \citep[LVG, e.g., ][]{1974ApJ...189..441G,1974ApJ...187L..67S} analysis. This model calculates the radiative transfer of molecular emission lines assuming a spherically symmetric cloud with a uniform photon escape probability and radial velocity gradient of $dv/dr$, where $dv$ is the half-width half maximum of CO line width and $dr$ is the radius of cloud C. We adopt $dv/dr$ = 1.2 km s$^{-1}$ / 1.2 pc $\sim1$ km s$^{-1}$ pc$^{-1}$, assuming the kinematic distance of 5.5 kpc. We also used the abundance ratio of [$^{12}$CO/H$_2$] = $5 \times 10^{-5}$ \citep{1987ApJ...315..621B} and [$^{12}$CO]/[$^{13}$CO] = 75 \citep{2004dimg.conf..253G}. Accordingly, we adopt $X/(dv/dr)$ as $5\times10^{-5}$ (km s$^{-1}$ pc$^{-1}$)$^{-1}$, where $X$ is the abundance ratio of [$^{12}$CO/H$_2$].

\begin{table*}[htbp]
  \begin{center}
    \caption{Properties of IRAS point sources and H{\sc ii} regions around Kes~79}
    \begin{tabular}{cccccccccc} \hline\hline 
    \shortstack{No. \\ (1)} & \shortstack{Source Name \\ (2)} & \shortstack{$l$ (deg) \\ (3)} & \shortstack{$b$ (deg) \\ (4)} & \shortstack{$F_\mathrm{12}$ \\ (Jy) \\ (5)} & \shortstack{$F_\mathrm{25}$ \\ (Jy) \\ (6)} & \shortstack{$F_\mathrm{60}$ \\ (Jy) \\ (7)} & \shortstack{$F_\mathrm{100}$ \\ (Jy) \\ (8)} &  \shortstack{Regions \\ (9)} & \shortstack{References \\ (10)}  \\ \hline 
    IR01 & IRAS~18501+0038 & 33.69 & \phantom{$-$}0.05 & 1.5 & 1.4 & 50.5 & 419 & B &  [1] \\
    IR02 & IRAS~18502+0034 & 33.66 & $-$0.02 & 2.5 & 20.4 & 158 & 419 & E & [1]  \\
    IR03 & IRAS~18501+0039 & 33.72 & \phantom{$-$}0.06 & 2.7 & 3.0 & 50.5 & 295 & B & [1]  \\
    IR04 & IRAS~18503+0041 & 33.77 & \phantom{$-$}0.03 &2.7 & 1.8 & 37.6 & 162 & & [1] \\
    IR05 & IRAS~18506+0038 & 33.75 & $-$0.07 & 2.1 & 10.2 & 82.0 & 122 & D &  [1] \\
    IR06 & IRAS~18496+0037 & 33.62 & \phantom{$-$}0.14 & 4.4 & 2.6 & 21.0 & 192 & & [1] \\
    IR07 & MSX5C G033.8580$-$00.0042 & 33.86 & $-$0.00 & & & & & & [2] \\
    IR08 & IRAS 18498+0045B & 33.76 & \phantom{$-$}0.16 & 5.3 & 3.4 & 16.4 & 162 & A & [1] \\
    IR09 & IRAS 18498+0028  & 33.52 & \phantom{$-$}0.02 & 4.1 & 11.3 & 72.9 & 176 & & [1] \\
    IR10 & IRAS 18496+0042 & 33.71 & \phantom{$-$}0.18 & 1.7 & 3.4 & 26.3 & 192 & A & [1]  \\
    IR11 & [GE91] GGD 30 IRS 8 & 33.59 & $-$0.14 & & & & & & [3] \\ 
    IR12 & IRAS 18498+0045A & 33.77 & \phantom{$-$}0.17 & 2.1 & 3.4 & 16.4 & 141 & A & [1] \\
    IR13 & IRAS 18494+0038 & 33.63 & \phantom{$-$}0.19 & 2.9 & 3.8 & 50.8 & 192 & & [1] \\
    IR14 & IRAS 18504+0025 & 33.54 & $-$0.13 & 2.6 & 5.9 & 98.1 & 305 & & [1] \\
    H{\sc ii}01 & [KC97c] G033.7+00.0 & 33.65 & \phantom{$-$}0.04 & & & & & B & [4] \\
    H{\sc ii}02 & HRDS G033.753$-$0.063 & 33.75 & $-$0.06 & & & & & D & [5] \\
    H{\sc ii}03 & MSX6C G033.5237+00.0198 & 33.52 & \phantom{$-$}0.02 & & & & & & [6] \\ \hline
    \end{tabular}
  \end{center}
\tablecomments{Column (1): numbers in Figure \ref{fig4}. Column (2): name of infrared point sources (IR01--IR14) and H{\sc ii} regions (H{\sc ii}01--H{\sc ii}03) around Kes~79. Columns (3)--(4): source positions in Galactic coordinate. Columns (5)--(8): fluxes of 12, 25, 60, and 100 $\mu$m. Column (9): region name defined in Section \ref{subsec:ratio}. Column (10): references,  [1] \cite{1988iras....1.....B}, [2] \cite{2001AAS...19913010E}, [3] \cite{1991A&A...241..589G}, [4] \cite{1997ApJ...488..224K}, [5] \cite{2012A&A...537A...1A}, [6] \cite{2003yCat.5114....0E}.}
 　 \label{table1}
% \vspace*{0.5cm}
\end{table*}

\begin{figure}
\begin{center}
\includegraphics[width=82mm,clip]{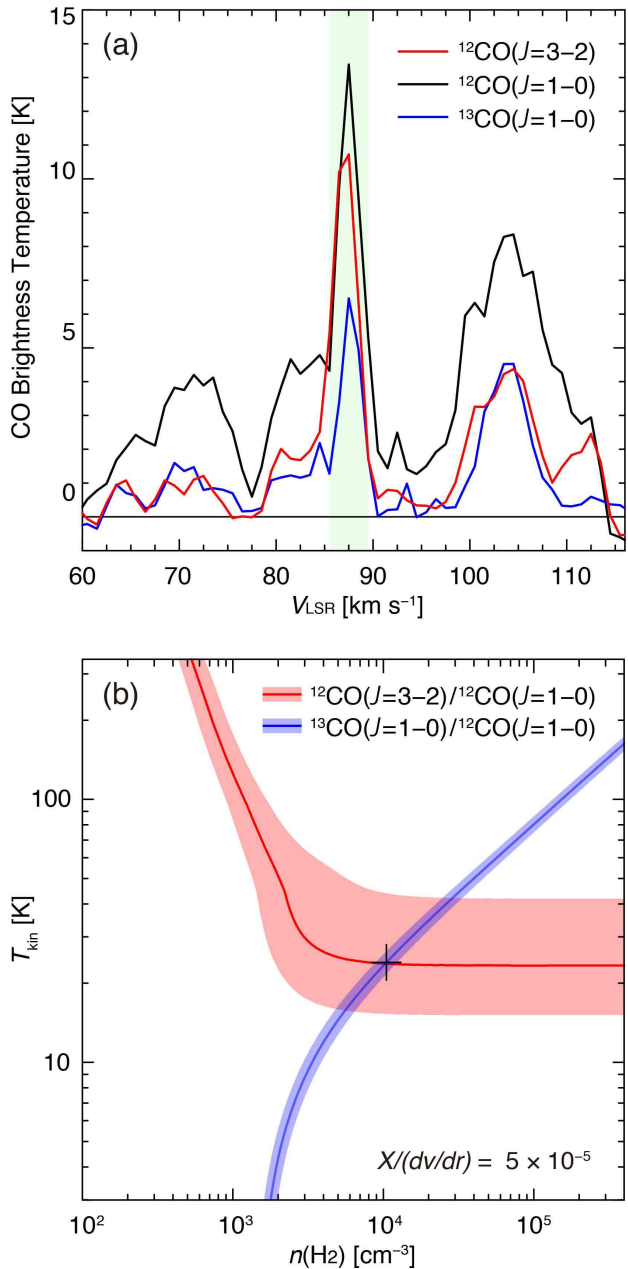}
\caption{(a) CO spectra toward cloud C. The black, red, and blue spectra represent $^{12}$CO($J$ = 1--0), $^{12}$CO($J$ = 3--2), and $^{13}$CO($J$ = 1--0) emission lines, respectively. (b) LVG results on the number density $n$(H$_2$) and kinematic temperature $T_\mathrm{kin}$ plane toward cloud C. The red and blue solid lines indicate the intensity ratios of $^{12}$CO($J$ = 3--2)/$^{12}$CO($J$ = 1--0) and $^{13}$CO($J$ = 1--0)/$^{12}$CO($J$ = 1--0), respectively. The cross represents the best fit values of $n$(H$_2$) and $T_\mathrm{kin}$ (for details, see the text).}
\vspace*{-0.5cm}
\label{fig5}
\end{center}
\end{figure}

Figure \ref{fig5}a shows the CO spectra toward cloud C. The velocity range used for the LVG analysis is shown shaded. Each spectrum was smoothed to much the FWHM of $^{12}$CO($J$ = 1--0) emission line. We obtained the intensity ratios of $R_\mathrm{3-2/1-0} \sim 0.84$ and $^{12}$CO $J$ = 1--0 /$^{13}$CO $J$ = 1--0 (hereafter $R_\mathrm{^{12}CO/^{13}CO}$) $\sim 0.45$.

Figure \ref{fig5}b shows the result of the LVG analysis of cloud C. The red and blue lines indicate $R_\mathrm{3-2/1-0}$ and $R_\mathrm{^{12}CO/^{13}CO}$, respectively. The errors (as shown in shaded areas in Figure \ref{fig5}b) are estimated with 1$\sigma$ noise levels for each spectrum, and relative calibration error of 5$\%$. Since $^{12}$CO($J$ = 1--0) and $^{13}$CO($J$ = 1--0) were simultaneously observed using the FOREST, the relative calibration error of 5$\%$ is canceled for the case of $R_\mathrm{^{12}CO/^{13}CO}$. We finally obtained the kinematic temperature $T_\mathrm{kin} \sim 24$ K (16--42 K) and the  number density $n$(H$_2$)$\sim 10000$ cm$^{-3}$ (6000--30000 cm$^{-3}$).

\section{DISCUSSION} \label{sec:discussion}
\subsection{Molecular and atomic clouds associated with the SNR Kes~79}\label{subsec:associated}
Previous studies of the interstellar gas toward Kes~79 suggest the 100 km s$^{-1}$ cloud is associated with the SNR \citep{1992MNRAS.254..686G,2016ApJ...816....1K,2016ApJ...831..192Z}. This claim is mainly based on two elements: (1) the 100 km s$^{-1}$ cloud shows good spatial correspondence with the X-ray/radio-continuum shell and filamentary structures, and (2) the broad molecular line with 20 km s$^{-1}$ width was possibly created by shock interactions. In this section, we show instead that the 80 km s$^{-1}$ cloud is the one most likely to be associated with the SNR Kes~79 rather than the 30 or 100 km s$^{-1}$ clouds.

First, we argue that it is difficult to decide which clouds are truly associated with the SNR Kes~79 from spatial comparisons alone. On the basis of our comparative study of the CO, H{\sc i}, radio-continuum, and X-rays data, the 30, 80, and 100 km s$^{-1}$ clouds all show good spatial correspondence with the X-ray filaments and the radio shell (see Section \ref{subsec:overview}). Spatially, the 30 km s$^{-1}$ cloud has a cavity-like structure in H{\sc i} that coincides with the radio shell (Figure \ref{fig2}d). The overall CO distribution of the 80 km s$^{-1}$ cloud tends to encircle the X-ray filaments and the radio shell (Figure \ref{fig2}b). In the 100 km s$^{-1}$ cloud, the dense molecular clouds appear to lie along the southeastern shell of the SNR (Figure \ref{fig2}c).

In contrast, the velocity structures of the clouds are more useful for distinguishing which are really associated with the SNR. Expanding shell-like structures of CO and H{\sc i} are seen in 80 km s$^{-1}$ cloud (see Figure \ref{fig3}b), whereas the 30 and 100 km s$^{-1}$ clouds have no characteristic structures in either the CO or H{\sc i} (see Figures \ref{fig3}a and \ref{fig3}c). The 4 km s$^{-1}$ expanding gas motions in the 80 km s$^{-1}$ clouds are thought to be created by SNR shockwaves and/or a strong stellar wind from the progenitor of the supernova explosion \citep[][]{1974ApJ...188..501C}. The typical expansion velocity of the molecular and atomic clouds associated with the SNR is to be $\sim7$--13 km s$^{-1}$ \citep[e.g.,][]{1989MNRAS.237..277L,2006ApJ...642..307Y,2012ApJ...746...82F,2013ApJ...774...10S,2017JHEAp..15....1S}. The expansion velocity of Kes~79 is found to be $\sim4$ km s$^{-1}$, which is similar to previous studies. 

The spatial distribution of $R_\mathrm{3-2/1-0}$ provides another source of information to help decide which molecular clouds are associated with the SNR. In the direction of Kes~79, we have identified some regions that have high intensity ratios of $R_\mathrm{3-2/1-0} > 0.8$ and have labeled them A--E. The positions of A, B, D, and E in the 30 and 100 km s$^{-1}$ clouds correspond to IRAS point sources and H{\sc ii} regions (see Figure \ref{fig4} and Table \ref{table1}), which may thus be the sources that have heated these molecular clouds. Furthermore, the existence of H{\sc ii} regions in the direction of Kes~79 does not exclude the possibility that the broad molecular line in the 100 km s$^{-1}$ cloud was created by star-formation activity \citep[e.g., by multiple outflows from young stellar objects;][]{2010ApJ...724...59S}. In contrast, region C in the 80 km s$^{-1}$ cloud has no extra heating sources (e.g., infrared sources or H{\sc ii} regions), and the kinematic temperature of the region C is $\sim 24$ K, indicating that shock heating has occurred \citep[e.g.,][]{2013ApJ...768..179Y}.

\begin{figure}
\begin{center}
\includegraphics[width=\linewidth,clip]{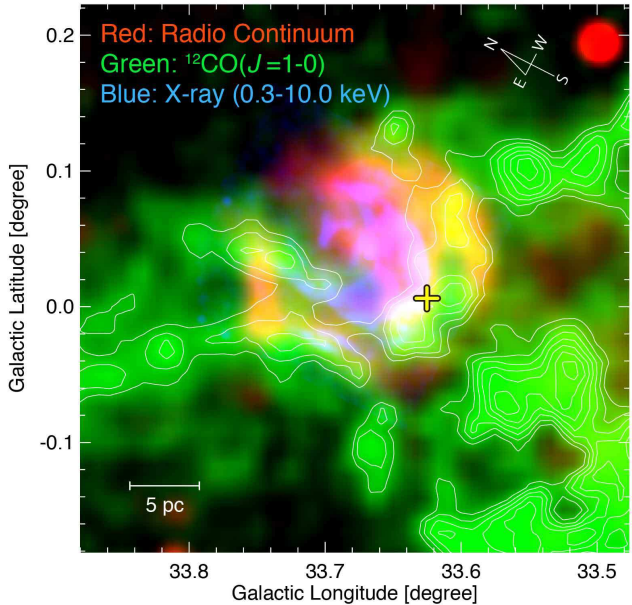}
\caption{RGB image of the SNR Kes~79. The red, green, and blue colors represent the intensity maps of radio continuum, $^{12}$CO($J$ = 1--0), and X-rays. The superposed contours indicate $^{12}$CO($J$ = 1--0) integrated intensity. The contour levels are every 3 K km s$^{-1}$ from 20 K km s$^{-1}$. The integration velocity range of CO is $V_\mathrm{LSR}$ = 82.5--86.5 km s$^{-1}$. The position of GeV gamma-ray peak is indicated by the yellow cross \citep{2014ApJ...783...32A}.}
\label{fig6}
\end{center}
\end{figure}

A detailed spatial comparison among the radio continuum, CO, X-rays, and gamma-rays provides further evidence for the association of the 80 km$^{-1}$ cloud with the SNR. Figure \ref{fig6} shows a three-color image of Kes~79. The molecular cloud at 80 km s$^{-1}$ delineates the outer boundary of the thermal X-ray shell and is embedded within the radio-continuum shell. We note that the radio-continuum shell is complementary to the two CO peaks in the southwest region, which is a typical signature of a shock-cloud interaction. According to \cite{2017JHEAp..15....1S}, interactions between shockwaves and inhomogeneous gas clumps enhance the synchrotron radio continuum around the gas clumps via amplifications of turbulence and magnetic fields \citep[c.f.,][]{2012ApJ...744...71I}. Although the angular resolution of $Fermi$ LAT is large, we find a good spatial correspondence between the peak positions of the molecular cloud and gamma-rays.

The cavity wall of dense gas surrounding the SNR is also consistent with the thermal X-ray properties of the SNR Kes~79. \cite{2016PASJ...68S...8S} discovered that the CIE plasma extends to the outer radio shell, whereas the NEI plasma is spatially concentrated within the inner radio shell. We can interpret the X-ray properties with two scenarios: (1) the CIE plasma in the shock front was produced by the interaction between the forward shock and the dense gas wall, and (2) the reflected shock generated by the shock-cloud interaction thermalized the supernova ejecta inside the radio shell, which is observed as the NEI plasma. Moreover, since the bremsstrahlung X-ray flux is proportional to the square of the ambient gas density, the bright thermal X-rays toward the cloud C gives a further support of the shock-cloud interaction. In light of these considerations, we conclude that the 80 km$^{-1}$ cloud is the one most likely to be associated with the SNR Kes~79.

\subsection{Distance and age}\label{subsec:distance}
On the basis of the physical association between the SNR shockwaves and the 80 km s$^{-1}$ cloud, we here discuss the age and distance of Kes~79. In Sections \ref{subsec:overview} and \ref{subsec:expanding}, we found that the velocity range of the 80 km s$^{-1}$ cloud extends from $\sim81.5$ to $\sim89.5$ km s$^{-1}$; this corresponds to a central radial velocity of $\sim85.5$ km s$^{-1}$ and an expansion velocity of $\sim4$ km s$^{-1}$. By adapting the Galactic rotation curve model of \cite{1993A&A...275...67B}, we therefore obtain the kinematic distance of Kes~79 as $5.5 \pm 0.3$ kpc for the near side and $8.7 \pm 0.3$ kpc for the far side.

\begin{deluxetable*}{lccccccc}[]
\tablewidth{\linewidth}
\tablecaption{Comparison of physical parameters in the $\gamma$-ray SNRs}
\tablehead{%& \multicolumn{3}{c}{Young SNRs}& & \multicolumn{2}{c}{Middle-Aged SNRs}\\
%\cline{2-4}\cline{6-7}
& RX~J1713.7$-$3946$^\mathrm{a)}$ & Vela Jr.$^\mathrm{b)}$ & HESS~J1731$-$347$^\mathrm{c)}$ & & Kes~79$^\mathrm{d)}$ & W44$^\mathrm{e)}$}
\startdata
Age (yr) & 1600 & 2400 & 4000 & & $8300 \pm 500$ & 20000\\
Distance (kpc) & 1 & 0.75 & 5.2 & &  5.5 & 3\\
Diameter (pc) & 17.4 & 11.8 & 22 & & 16 & 25\\
Molecular proton mass (10$^4$ $M_{\odot}$) & 0.9  & 0.1 & 5.1 && 1.7 & 40\\
Atomic proton mass (10$^4$ $M_{\odot}$) & 1.1  & 2.5 & 1.3 && 0.2 & 6 \\
Total proton mass (10$^4$ $M_{\odot}$) & 2.0  & 2.6 & 6.4 && 1.9 & 46 \\
$N_\mathrm{p}$(H$_2$) (cm$^{-3}$) & 60  & 4 & 64 && 310 & 180 \\
$N_\mathrm{p}$(H{\sc i}) (cm$^{-3}$) & 70  & 96 & 16 && 50 & 20 \\
$N_\mathrm{p}$(H$_2$ + H{\sc i}) (cm$^{-3}$) & 130  & 100 & 80 && 360 & 200 \\
$N_\mathrm{p}$(H$_2$)/$N_\mathrm{p}$(H{\sc i}) & 0.9 & 0.04 & 4 && 6 & 9\\ 
Total CR proton energy ($10^{48}$ erg) & 0.4$^\dagger$ & 0.7 & 5 && 5 & 10 \\ 
\enddata
\tablecomments{\\a) \cite{2012ApJ...746...82F}.\\b) \cite{2017ApJ...850...71F}.\\c) \cite{2014ApJ...788...94F}.\\d) This work.\\e) \cite{2013ApJ...768..179Y}.\\$^\dagger$ We adopt $N_\mathrm{p}$(H$_2$ + H{\sc i}) to the latest hadronic model presented by \cite{2018AA...612A...6H}.}
%\vspace*{-0.5cm}
\label{table2}
\end{deluxetable*}

It is difficult to distinguish the near and far sides if the two distances are relatively close to each other, but Kes~79 is likely to be located at the near side. Because the absorbing column density toward Kes~79 \citep[$\sim1.5 \times 10^{22}$ cm$^{-2}$,][]{2016PASJ...68S...8S} is consistent with the column density of foreground interstellar gas between the Sun and the SNR if it is located at the near side. The foreground interstellar gas on the near side corresponds to the velocity range from 2 to 81.5 km s$^{-1}$. The column density of foreground interstellar gas is $N_\mathrm{p}$(H$_2$ + H{\sc i}) = $2 \times N$(H$_2$) + $N$(H{\sc i}), where $N$(H$_2$) is the column density of molecular hydrogen and $N$(H{\sc i}) is the column density of atomic hydrogen. In this work, we use the relations $N$(H$_2$) = $1.0 \times 10^{20} \times W$(CO) cm$^{-2}$ \citep{2017ApJ...838..132O} and $N$(H{\sc i}) = $1.823 \times 10^{20} \times W$(H{\sc i}) cm$^{-2}$ \citep{1990ARA&A..28..215D}, where $W$(CO) and $W$(H{\sc i}) are the integrated intensities of CO and H{\sc i}, respectively. We thus obtain $N_\mathrm{p}$(H$_2$ + H{\sc i}) = $\sim3 \times 10^{22}$ cm$^{-2}$, which is enough to explain the absorbing column density toward Kes~79. On the other hand, $N_\mathrm{p}$(H$_2$ + H{\sc i}) for the far side distance is estimated to be $\sim6 \times 10^{22}$ cm$^{-2}$ assuming the velocity range from 2 to 110 km s$^{-1}$, which is too large as compared with the absorbing column density. It is therefore that the near side distance is more reasonable than the far side distance. To distinguish more accurately between the near and far sides, we need H{\sc i} absorption studies using H{\sc i} data with a finer angular resolution because of the strong radio continuum from Kes~79. We hereafter use the near-side distance of $5.5 \pm 0.3$ kpc. 

Because the shockwave from Kes~79 interacts strongly with the 80 km s$^{-1}$ cloud, we can estimate the dynamical age of the SNR assuming that it is in the Sedov-Taylor phase \citep{1959sdmm.book.....S}. The dynamical age $t_{\rm age}$ of the SNR is then given by;
\begin{eqnarray}
t_\mathrm{age} = \frac{2 R_\mathrm{sh}}{5 V_\mathrm{sh}}%2 R_\mathrm{sh} / 5 V_\mathrm{sh}%
\label{eq1}
\end{eqnarray}
where $R_\mathrm{sh}$ is the radius of the SNR, and $V_\mathrm{sh}$ is the shock velocity. We adopt the $R_\mathrm{sh} \sim 8$ pc from the 5 arcmin radius of the outer radio shell. According to \cite{2016PASJ...68S...8S}, the best-fit CIE plasma temperature of $T_\mathrm{e}$ is $\sim0.17 \pm 0.02$ eV. Assuming the electron--ion temperature equilibration, $V_\mathrm{sh}$ can be described as $[(16 k_\mathrm{B} T_\mathrm{e})/(3 \mu m_\mathrm{p})]^{0.5}$, where $k_\mathrm{B}$ is is Boltzmann's constant, $\mu = 0.604$ is mean atomic weight, and $m_\mathrm{p}$ is the atomic hydrogen mass. We therefore obtain $V_\mathrm{sh} = 380 \pm 20$ km s$^{-1}$ and the dynamical age $t_\mathrm{age} = 8300 \pm 500$ yr, indicating that the SNR Kes~79 can be categorized as a middle-aged SNR.

\subsection{Total CR Protons Energy}\label{subsec:cr}
\cite{2014ApJ...783...32A} discovered a GeV gamma-ray excess in the direction of Kes~79, which is consistent with hadronic gamma rays. To obtain the total CR proton energy, we first need to determine the total interstellar proton density associated with the SNR. The total mass of interstellar gas associated with Kes~79 is $\sim1.9 \times 10^4$ $M_{\odot}$ within a radius of 8 pc, and the mass of the molecular component is $\sim1.7 \times 10^4$ $M_{\odot}$ and that of the atomic component is $\sim0.2 \times 10^4$ $M_{\odot}$. Here, we adopt the helium abundance of the molecular cloud to be $\sim20$ $\%$ and assume the H{\sc i} to be optically thin. Adopting a shell thickness of $\sim 3$ pc, we find the total interstellar proton density to be $\sim360$ cm$^{-3}$, with the proton density of the molecular component being $\sim310$ cm$^{-3}$, and that of the atomic component being $\sim50$ cm$^{-3}$. Here, we used the relation between $N$(H$_2$) and $W$(CO) for estimating the molecular component; $N$(H{\sc i}) and $W$(H{\sc i}) for estimating the atomic component.

The total CR proton energy $W_\mathrm{pp}^\mathrm{tot}$ can then be derived by using the equation \citep[e.g.,][]{2006A&A...449..223A};
\begin{eqnarray}
W_\mathrm{pp}^\mathrm{tot} \sim t_\mathrm{pp \rightarrow \pi_0} \times L_\gamma 
\label{eq2}
\end{eqnarray}
where $t_\mathrm{pp \rightarrow \pi_0} \sim 4.5 \times 10^{15}$ ($n$/1 cm$^{-3}$)$^{-1}$ is the characteristic cooling time of the protons and $L_\gamma$ is the gamma-ray luminosity. According to \cite{2014ApJ...783...32A}, $L_\gamma$(0.1--100 GeV) $\sim3.8 \times 10^{35}$ ($d$/5.5 kpc)$^{2}$ erg s$^{-1}$, where $d$ is the distance to Kes~79. This values is consistent with a relation between the SNR radius and gamma-ray luminosity \citep{2016PASJ...68S...5B}. The resulting value of $W_\mathrm{pp}^\mathrm{tot}$(1--1000 GeV) is $\sim 1.7 \times 10^{51} (d$/5.5 kpc)$^2$ ($n$/1 cm$^{-3})^{-1}$. Adopting the $n \sim 360$ cm$^{-3}$ and $d \sim 5.5$ kpc, we finally obtain $W_\mathrm{pp}^\mathrm{tot}$(1--1000 GeV) $\sim 5 \times 10^{48}$ erg, corresponding to $\sim0.5$ $\%$ of the typical $\sim10^{51}$ erg kinematic energy of supernova explosion. 

A comparison between young gamma ray SNRs (RX~J1713.7$-$3946, Vela Jr., and HESS~J1731$-$347) and middle-aged SNRs (Kes~79 and W44) is given in Table \ref{table2}. The total interstellar proton masses are not much different among the SNRs, except for W44, but the abundance ratios of molecular and atomic protons $N_\mathrm{p}$(H$_2$)/$N_\mathrm{p}$(H{\sc i}) are slightly different. We note that both the total CR proton energy and the ratio $N_\mathrm{p}$(H$_2$)/$N_\mathrm{p}$(H{\sc i}) increase as an SNR ages. We can easily interpret the total CR proton energy as indicating that the efficiency of CR acceleration may increase in time, as has been previously suggested in some pioneering studies \citep[e.g.,][]{2012ApJ...746...82F,2013ApJ...768..179Y,2014ApJ...788...94F}. As for the ratio $N_\mathrm{p}$(H$_2$)/$N_\mathrm{p}$(H{\sc i}), we propose the hypothesis that the degree of shock propagation into a dense molecular cloud may affect the time variation of $N_\mathrm{p}$(H$_2$)/$N_\mathrm{p}$(H{\sc i}). Further numerical simulations with three-dimensional magnetohydrodynamics will test this in detail.

To obtain a complete understanding of the total CR protons energy, the role of relativistic particle escape in this system should be considered. According to \cite{2012ApJ...749L..35U}, the contribution of escaped CR protons in W44 is estimated to be $\sim0.3$--$3\times10^{50}$ erg. Kes~79 also has a potential to detect the gamma-rays originated from the escaped CR protons and the total CR proton energy will be increased. Moreover, re-acceleration of pre-existing CR protons by radiative shocks might be an important component to our understanding of the relativistic particle populations in this remnant \citep[e.g.,][]{2010ApJ...723L.122U}. Further gamma-ray observations with high-spatial resolution and high-sensitivity using the Cherenkov Telescope Array (CTA) will allow us to accurately estimate the total CR proton energy of Kes~79.

\section{SUMMARY} \label{sec:summary}
We have presented a new combined analysis of both the CO and H{\sc i} toward the gamma ray SNR Kes~79. We summarize the findings below:

\begin{enumerate}
\item We found three candidates for the molecular/atomic clouds associated with the SNR Kes~79, which lie at radial velocities of $\sim30$, 80, and 100 km s$^{-1}$. These clouds show good spatial correspondence with the radio and X-ray shells.
\item Expanding gas motions, with $\Delta V \sim 4$ km s$^{-1}$, were obtained in the 80 km s$^{-1}$ cloud, which were likely created by the SNR shockwaves and/or a strong stellar wind from the progenitor of the supernova explosion. Furthermore, the southern CO clump of the 80 km s$^{-1}$ cloud--named cloud C--also showed a high-intensity ratio $> 0.8$ for CO 3--2/1--0, indicating a kinematic temperature of $\sim24$ K owing to heating by the shock interaction. We conclude that the 80 km s$^{-1}$ cloud is the one most likely to be associated with the SNR Kes~79.
\item The kinematic distance of $\sim5.5$ kpc and the dynamical age of $8300 \pm 500$ yr of Kes~79 were updated using the Galactic rotation curve model of \cite{1993A&A...275...67B} and Sedov's self-similar solution \citep{1959sdmm.book.....S}. The near-side distance of $\sim5.5$ kpc is consistent with the absorbing column density derived by X-ray spectroscopy and the foreground gas density derived from the CO/H{\sc i} datasets.
\item We obtained a total CR proton energy of $W_\mathrm{pp}^\mathrm{tot}$(1-- 1000 GeV) $\sim 5 \times 10^{48}$ erg, which corresponds to 0.5 $\%$ of the typical $\sim10^{51}$ erg kinematic energy of a supernova explosion. We note that both the total CR proton energy $W_\mathrm{pp}^\mathrm{tot}$ and the abundance ratios of molecular and atomic protons $N_\mathrm{p}$(H$_2$)/$N_\mathrm{p}$(H{\sc i}) increase as the SNR ages. We propose the hypothesis that the degree of shock propagation into a dense molecular cloud may affect both the time variation of $N_\mathrm{p}$(H$_2$)/$N_\mathrm{p}$(H{\sc i}) and the efficiency of CR acceleration, both of which may have increased in time.
\end{enumerate}

\acknowledgments
{\footnotesize{The Nobeyama 45-m radio telescope is operated by Nobeyama Radio Observatory, a branch of National Astronomical Observatory of Japan. The James Clerk Maxwell Telescope is operated by the East Asian Observatory on behalf of The National Astronomical Observatory of Japan, Academia Sinica Institute of Astronomy and Astrophysics, the Korea Astronomy and Space Science Institute, the National Astronomical Observatories of China and the Chinese Academy of Sciences (Grant No. XDB09000000), with additional funding support from the Science and Technology Facilities Council of the United Kingdom and participating universities in the United Kingdom and Canada. This work was financially supported by Grants-in-Aid for Scientific Research (KAKENHI) of the Japanese society for the Promotion of Science (JSPS, Grant Nos. 15H05694, 15K05107, 16H03961, and 16K17664). This work was also supported by “Building of Consortia for the Development of Human Resources in Science and Technology” of Ministry of Education, Culture, Sports, Science and Technology (MEXT, Grant No. 01-M1-0305). We are grateful to Dr. Ping Zhou and the anonymous referee for useful comments, which helped the authors to improve the paper.}}
\software{NOSTAR, CIAO \citep[v 4.8.1:][]{2006SPIE.6270E..1VF}}

%%%%%%%%%%%%%%%%%
%%% References %%%
%%%%%%%%%%%%%%%%%

\end{document}